# MSRyzen: An Agentic Platform for Unified Proteomics Analysis and Evidence-Centric Visualization

**Yucheng Liao[1,2,†], Jubao Cheng[2,†], Zexu Xiao[2], Tao Yi[2], Jiaxiang Ding[2], Weijie Zhang[2,3,*]**

## Affiliations

[1] **Center for Machine Learning Research, Peking University, Beijing, China 100871**

[2] **State Key Laboratory of Medical Proteomics, AI for Science Institute, Beijing, China 102206**

[3] **DP Technology, Beijing, China 100089**

## Contributions

W.Z. supervised the study. W.Z. and Y.L. conceived and designed the agent system. J.C. implemented the agentic system framework. Z.X. developed the visualization module. T.Y. collected and prepared the datasets for evaluation. J.D. helped perform the evaluation. Y.L., J.C., and W.Z. drafted and revised the manuscript. All authors reviewed and approved the final manuscript.

## *Corresponding author

Correspondence to: zhangwj@aisi.ac.cn; Y.L. and J.C. contributed equally to this work.

**Abstract**

Mass spectrometry (MS)-based proteomics is essential for comprehensive protein characterization, yet it relies on multistep computational workflows that typically demand tedious manual parameter tuning and complex software integration. Here, we present MSRyzen, an agent-driven framework that provides automated parameter optimization and extensible tool integration across bottom-up and top-down proteomics workflows. MSRyzen translates natural-language prompts into directed acyclic workflow graphs, optimizes analytical parameters guided by quantitative metrics and user-defined objectives, and incorporates new third-party software through standardized callable interfaces. Across four public top-down datasets, automated parameter optimization increased proteoform identifications by 3.6% to 15.8% compared to default settings at a 1% false discovery rate. To ensure transparent data interpretation, an interactive visualization interface coordinates identification metrics with peak-level spectral evidence and sequence coverage. Integrating adaptive tuning with multi-tiered spectral inspection, MSRyzen delivers an end-to-end, fully automated framework, dramatically reducing the operational barrier to modern mass spectrometry data analysis.

**Introduction**

Driven by continuous advancements in high-resolution mass spectrometry and analytical methodologies, mass spectrometry (MS)-based proteomics has become indispensable for deciphering complex biological systems and disease mechanisms [1][2][3]. To achieve comprehensive structural and functional profiling, research has primarily evolved along two complementary strategies: bottom-up peptide profiling and top-down intact proteoform characterization [4][5][6][7]. Despite their distinct analytical strengths, transforming raw spectral data into accurate and efficient identifications remains a persistent rate-limiting bottleneck for both approaches[8][9].

Standard MS data interpretation involves a series of interdependent steps, including spectral deconvolution [10][11][12], feature detection [13][14], database searching [15][16], post-translational modification (PTM) localization [17][18], rescoring [19][20], and false discovery rate (FDR) control [21]. While specialized algorithms have been developed for individual steps, they lack end-to-end processing capabilities. To address these limitations, researchers have developed integrated software platforms such as FragPipe [15], AlphaPeptDeep [22] and MaxQuant [13] for bottom-up proteomics, as well as MASH Native [23] and TDEase [24] for top-down proteomics. Nevertheless, these platforms largely operate as isolated ecosystems. Executing end-to-end workflows still demands manual integration of disparate command-line tools and empirical tuning of interdependent parameters. In addition, suboptimal parameter configurations often introduce variability in sensitivity and FDR control, making analytical outcomes highly operator-dependent and hindering high-throughput implementation.

To bridge these gaps, we present MSRyzen, an agent-driven framework designed for automated parameter optimization and extensible software orchestration across bottom-up and top-down proteomics. MSRyzen compiles natural-language instructions into executable workflow graphs and utilizes intermediate identification statistics alongside quality metrics to drive iterative parameter refinement. Standardized adapter interfaces enable dynamic registration of third-party analytical tools based on structured metadata and software documentation. Furthermore, an integrated web interface couples analytical outputs directly to primary sequence coverage, chromatographic traces, and annotated fragment-ion spectra for visual validation. We demonstrate the utility of MSRyzen through automated top-down parameter optimization, harmonized cross-engine comparison of bottom-up results across FragPipe, MaxQuant, and Comet [25], and joint profiling linking bottom-up peptide discovery with intact-MS1 screening in a cardiomyocyte maturation study. By combining autonomous parameter optimization, cross-paradigm bottom-up and top-down integration, and highly extensible tool orchestration, MSRyzen bridges complex multi-software workflows with transparent, inspectable spectral evidence in a single cohesive environment.

## Experimental Section

### Overall System Architecture.

MSRyzen integrates an agent-driven workflow engine, an execution runtime, and a web visualization platform. Upon receiving natural-language prompts, MS datasets, and a FASTA database, the system constructs a shared task state and compiles an execution graph structured as a directed acyclic graph (DAG). Within this DAG, nodes represent discrete analytical operations and their associated parameters, while edges encode data dependencies between tasks. Following execution, analytical results, runtime status, and quality-control metrics are fed back to the agent, facilitating automated parameter refinement and downstream analyses (Figure 1, S1).

### Agent-based Workflow Implementation.

The agent workflow was implemented using LangGraph, with model access managed through a LiteLLM router supporting OpenAI-compatible endpoints. A centralized task-state object maintains the global execution context, including DAG topology, tool parameters, runtime status, file paths, error logs, and quality-control metrics. Prior to dispatch, candidate workflows undergo static validation to detect circular dependencies and ensure analytical tool compatibility, followed by user confirmation. An asynchronous scheduler then resolves execution order and maps runnable nodes to corresponding command-line processes.For top-down proteomics, the framework integrates MSConvert [28] followed by TopFD/FLASHDeconv [11][29] and TopPIC [16], as well as the PbfGen–ProMex–MSPathFinderT pipeline [26]. For bottom-up workflows, core FragPipe modules (including MSFragger [15], MSBooster [20], and IonQuant [30]) are encapsulated as modular, schedulable operators that assemble into customized DAG pipelines. Finally, dedicated parsers ingest process outputs back into the task state, while node-level checkpointing enables fault recovery and selective re-execution without recomputing upstream tasks.

### Automatic Parameter Optimization Implementation.

Parameter optimization is abstracted into an automated feedback loop governed by user-defined quantitative objectives, defaulting to the maximization of confident identification counts under stringent false discovery rate and quality thresholds. Tool-output parsers convert raw results into structured performance records, like the number of proteoforms identified at 1% FDR by TopPIC and an inventory of unexplained mass shifts. Guided by these diagnostic profiles, the agent iteratively refines tool configurations to optimize search sensitivity. To maintain high computational efficiency during parameter benchmarking, the framework selectively reevaluates only the affected downstream subgraphs within the directed acyclic graph. Modifying a search parameter, for instance, restricts re-execution solely to the TopPIC search and subsequent downstream steps while reusing cached TopFD deconvolution outputs, avoiding redundant upstream computation. Each iteration logs parameters, artifact paths, and resulting metrics in the shared state. Candidate configurations are ranked against the selected objective function, and the highest-performing valid set is retained. The tuning loop halts when convergence criteria or predefined iteration limits are met.

**Extensible Software Integration.**

Analytical tools are encapsulated within callable interfaces specifying their operational functions, I/O schemas, and parameters. New tools are registered dynamically by the agent using CLI help outputs or API documentation paired with a structured definition file (tool.md). Tool-specific wrappers map workflow configurations to executable commands and register generated artifacts for downstream DAG nodes, enabling seamless integration into the scheduler without altering the underlying graph representation.

**Visualization System Implementation.**

To support versatile spectral inspection, the visualization platform accommodates diverse mass spectrometry workflows across bottom-up and top-down regimes, encompassing data-dependent acquisition (DDA), data-independent acquisition (DIA), and standalone raw or mzML files. The viewer directly parses outputs from varied processing pipelines, including DIA-NN and MaxQuant, mapping identification records to primary sequence coverage, precursor isotope envelopes, and annotated tandem spectra. For nonstandard analytical outputs, an underlying agent autonomously interprets data schemas, resolves ingestion errors, and registers verified parsing strategies for automated reuse. To ensure responsive rendering across large-scale mass spectrometry datasets, the backend converts raw spectral arrays and analysis results into an an indexed binary representation containing retention times, m/z values, intensities, chromatographic traces, and spectrum matches. PostgreSQL and SQLAlchemy manage metadata and relationships among datasets, proteins, proteoforms, peptides, and spectra. Spectral arrays are stored outside the relational database and retrieved through the spectral indexes. REST endpoints provide access to metadata, extracted-ion chromatograms, total ion chromatograms, and MS1/MS2 spectra. The frontend connects identification records to sequence annotations, precursor isotope envelopes, matched fragment ions, and mass-error plots. The indexed format and its data structures are described in the Figure S2.

**Datasets and Data Analysis.**

Four public top-down datasets (PXD019247 [31], PXD026123 [32], PXD046651 [33], and PXD029703 [34]) were used to compare default and optimized configurations at 1% FDR thresholds. Bottom-up data from PXD060431 [35] were analyzed in parallel using FragPipe, MaxQuant, and Comet, with outputs compared through protein-level set intersections and differential-abundance summaries. Cardiomyocyte maturation data from MSV000082985 [36] were analyzed as distinct bottom-up and intact-MS1 sample sets.

**Result and Discussion**

**Autonomous Parameter Optimization in Top-Down Proteomics**

Top-down proteomics searches typically require sample-specific and experiment-specific parameter adjustments, such as mass tolerances and signal-to-noise thresholds. To eliminate this bottleneck, MSRyzen performs autonomous parameter optimization and space exploration by iteratively proposing candidate configurations and evaluating against spectral identification performance (Figure 2a). We benchmarked this optimization capability across four top-down proteomics datasets. Relative to default search parameters, MSRyzen achieved a 3.6% to 15.8% increase in identified proteoforms across all cohorts (Figure 2b). Through the sequential optimization trajectory (Figure 2c), the agent drove proteoform recoveries from 340 to 518 by systematically calibrating mass tolerances, curbing PTM search penalties, and fine-tuning deconvolution cutoffs (MS1 S/N 1.5 and EnvCNN 0.2) to capture genuine low-abundance isotopic envelopes without violating the 1% false discovery rate threshold. In an E. coli case study (ColF18_Exp22_02, PXD019247), the optimized parameters identified 955 proteoforms and 260 proteins, compared to 672 proteoforms and 238 proteins under default settings, with 520 proteoforms and 227 proteins shared between configurations (Figure 2d). Spectral evidence inspection, including fragment-ion assignments and residual mass-error distributions, verified high-confidence identifications for both shared and configuration-specific proteoforms (Figure 2f). These results demonstrate that autonomous parameter exploration markedly improves proteoform recovery while maintaining stringent spectral reliability.

**Automated Software Integration and Multi-Tool Proteomic Analysis.**

To establish a future-ready architecture capable of seamlessly assimilating newly emerging or previously unencountered computational tools, we evaluated the open-ended integration framework by executing an end-to-end analysis of dataset PXD060431 across FragPipe alongside dynamically onboarded MaxQuant and Comet engines. To evaluate the onboarding of arbitrary third-party software, MSRyzen incorporated MaxQuant and Comet as entirely unseen components, automatically generating callable interfaces directly from raw help documentation and structured tool metadata without hardcoded wrappers (Figure 3a). The three outputs were integrated for protein-level comparison and downstream differential-abundance analysis. Together, they identified 2,284 proteins, including 1,474 shared across all three workflows, revealing a common identification set alongside software-specific detections (Figure 3b).

The differential-abundance results were then summarized using mean $\log_2$ fold changes and Fisher-combined *p*-values, while retaining whether each protein was significant in one, two, or all three analyses (Figure 3c). This allowed candidates to be assessed according to both their abundance changes and the consistency of statistical significance across software outputs. ACTC1, for example, was significant in all three analyses, with a Fisher-combined *p*-value of $4.8 \times 10^{-11}$. Crucially, these harmonized outputs are directly compatible with downstream bioinformatics skills native to the agent, facilitating automated pathway enrichment and functional annotation without manual data conversion (Figure 3d).

Beyond statistical prioritization, MSRyzen coordinates an automated dual-agent visualization framework to generate inspectable evidence views without manual intervention (Figure 3e). For ACTC1, the visualization agent autonomously synthesized a multi-tiered evidence profile, integrating reconciled sequence coverage across FragPipe, MaxQuant, and Comet alongside the extracted ion chromatogram and annotated MS/MS fragmentation spectrum of its representative doubly charged peptide AGFAGDDAPR (Figure 3f). This case illustrates how MSRyzen establishes a complete, closed-loop workflow connecting multi-tool identification, comparative bioinformatics, and autonomous spectral-level verification.

**Complementary Top-Down and Bottom-Up Analysis of Cardiomyocyte Maturation**

Integrating bottom-up (BU) and top-down (TD) proteomics provides complementary resolution across global protein abundance and intact proteoform diversity. However, cross-platform joint analysis has traditionally relied on laborious, manual cross-referencing between disparate workflows. To overcome this limitation, we used MSRyzen to orchestrate an automated cross-modality pipeline on the public human pluripotent stem cell (MassIVE accession MSV000082985).

At the global bottom-up level, analysis revealed extensive protein-group overlap between Day 30 and Day 59, with 2,916 shared protein groups (Figure 4a). Differential expression profiling identified 71 upregulated and 83 downregulated protein groups at Day 59 under nominal p-value and fold-change thresholds (Figure 4b).

Cross-referencing these candidates against established cardiac maturation markers (Table S1, 2) prioritized 11 key biomarkers for downstream targeted proteoform validation in the corresponding top-down intact-MS1 data.

Among these candidates, three biomarkers were robustly detected at the intact-protein level, with MLC2v exhibiting significant differential abundance (p = 6.8e-05) (Figure 4c, d). In-depth proteoform profiling further resolved distinct isoform and post-translational modification (PTM) dynamics, notably between atrial and ventricular regulatory light chains (MLC2a and MLC2v), as well as monophosphorylated MLC2v (+79.966 Da). During maturation from Day 31 to Day 60, total MLC2v signals increased whereas MLC2a signals decreased. Concurrently, the monophosphorylated MLC2v proteoform was sharply depleted and became virtually undetectable at Day 60 with a trend corroborated by representative isotopic MS1 spectral inspections (Figure 4e).

Beyond targeted tracking, MSRyzen conducted an agent-driven open proteoform search, which recovered the +248 Da mass shift on slow skeletal troponin I (ssTnI/TNNI1) originally documented in the reference study. By automatically querying the Unimod database, the agent matched the measured intact mass shift to a putative combinatorial Hex + Malonyl modification, with a mass deviation of only 0.0288 Da (all candidate annotations are detailed in Table S3). Together, these results demonstrate how MSRyzen seamlessly unifies high-throughput bottom-up candidate discovery, targeted top-down proteoform validation, and agent-driven open PTM characterization.

### Evidence-Centric Spectral Visualization.

To verify that the additional identifications obtained via parameter optimization were underpinned by robust spectral evidence, we investigated a proteoform of 50S ribosomal protein L33 (RL33_SALTY) uniquely resolved in the optimized workflow. The visualization interface opens with a global summary of identification metrics and confidence thresholds (Figure S3a), followed by a ranked table of identified proteoforms (Figure S3b). The selected proteoform was supported by 14 proteoform-spectrum matches (PrSMs) with a top E-value of $1.92 \times 10{-}17$ (Figure S3c), offering multiple orthogonal observations for spectral validation.

For individual PrSMs, the module coordinates primary sequence coverage with matched precursor and MS2 spectral profiles. The representative PrSM highlighted in Figure S3d exhibits a precursor m/z of 626.0609 (z = 10+), corresponding to a calculated intact proteoform mass of 6250.5765 Da across the 55-residue sequence and an N-terminal mass shift of -117.0433 Da. An interactive sequence map correlates detected fragment ions with their respective backbone cleavage sites, while synchronized MS1 and tandem mass spectra permit direct inspection of the precursor isotopic envelope alongside assigned fragment ions.

Spectral annotations depict the distribution and sequence coverage of assigned N- and C-terminal fragment ions across the spectrum (Figure S3e). At single-peak resolution, an integrated summary table delineates the observed m/z, relative intensity, charge state, sequence coordinate, and mass error for each annotation (Figure S3f). By providing multi-tiered, inspectable spectral evidence, the visualization module confirms that gains in identification yield stem from authentic, high-quality fragmentation events rather than relaxed statistical thresholds.

## Conclusions

In this work, we presented MSRyzen, an autonomous agentic framework that bridges high-level analytical intent with reliable computational execution across bottom-up and top-down mass spectrometry-based proteomics. MSRyzen unifies automated parameter optimization, extensible software orchestration, and spectral visualization into a modular computational architecture for proteomics. Across several representative scenarios, we demonstrated the progressive capabilities of the agent at distinct analytical levels. In top-down workflows, automated parameter optimization significantly enhanced proteoform recovery across diverse benchmarks while maintaining strict statistical cutoffs. In bottom-up analyses, the dynamic registration of external search engines, such as MaxQuant and Comet, supported direct, multi-engine comparative analysis without altering core workflow representations. The cardiomyocyte maturation study exemplified how global bottom-up discovery can inform targeted intact-MS1 profiling, providing candidate proteoform assignments for subsequent structural characterization. By bridging agent-based workflow planning with inspectable, peak-level evidence, MSRyzen minimizes the barrier to fine-tuning and expanding established proteomics tools, offering a scalable foundation for automated mass spectrometry data processing.

## Software Availability

MSRyzen is accessible through a web-based interface at http://ygit1484630.bohrium.tech:50003/, enabling users to interact with the agent and execute proteomics analysis workflows.

## Acknowledgement

This work was supported by the National Key R&D Program of China (2025YFA1309300).

## REFERENCES

[1] Aebersold, R., & Mann, M. (2003). Mass spectrometry-based proteomics. *Nature*, *422*(6928), 198-207.

[2] Aebersold, R., & Mann, M. (2016). Mass-spectrometric exploration of proteome structure and function. *Nature*, *537*(7620), 347-355.

[3] Guo, T., Steen, J. A., & Mann, M. (2025). Mass-spectrometry-based proteomics: from single cells to clinical applications. *Nature*, *638*(8052), 901-911.

[4] Steen, H., & Mann, M. (2004). The ABC's (and XYZ's) of peptide sequencing. *Nature reviews Molecular cell biology*, *5*(9), 699-711.

[5] Smith, L. M., & Kelleher, N. L. (2013). Proteoform: a single term describing protein complexity. *Nature methods*, *10*(3), 186-187.

[6] Roberts, D. S., Loo, J. A., Tsybin, Y. O., Liu, X., Wu, S., Chamot-Rooke, J., ... & Ge, Y. (2024). Top-down proteomics. *Nature Reviews Methods Primers*, *4*(1), 38.

[7] Smith, L. M., Agar, J. N., Chamot-Rooke, J., Danis, P. O., Ge, Y., Loo, J. A., ... & Consortium for Top-Down Proteomics. (2021). The human proteoform project: defining the human proteome. *Science advances*, *7*(46), eabk0734.

[8] Nesvizhskii, A. I. (2010). A survey of computational methods and error rate estimation procedures for peptide and protein identification in shotgun proteomics. *Journal of proteomics*, *73*(11), 2092-2123.

[9] Strauss, M. T., Bludau, I., Zeng, W. F., Voytik, E., Ammar, C., Schessner, J. P., ... & Mann, M. (2024). AlphaPept: a modern and open framework for MS-based proteomics. *Nature Communications*, *15*(1), 2168.

[10] Jeong, K., Kim, J., Gaikwad, M., Hidayah, S. N., Heikaus, L., Schlüter, H., & Kohlbacher, O. (2020). FLASHDeconv: ultrafast, high-quality feature deconvolution for top-down proteomics. *Cell Systems*, *10*(2), 213-218.

[11] Basharat, A. R., Zang, Y., Sun, L., & Liu, X. (2023). TopFD: A proteoform feature detection tool for top–down proteomics. *Analytical Chemistry*, *95*(21), 8189.

[12] Marty, M. T., Baldwin, A. J., Marklund, E. G., Hochberg, G. K., Benesch, J. L., & Robinson, C. V. (2015). Bayesian deconvolution of mass and ion mobility spectra: from binary interactions to polydisperse ensembles. *Analytical chemistry*, *87*(8), 4370.

[13] Cox, J., & Mann, M. (2008). MaxQuant enables high peptide identification rates, individualized ppb-range mass accuracies and proteome-wide protein quantification. *Nature biotechnology*, *26*(12), 1367-1372.

[14] Röst, H. L., Sachsenberg, T., Aiche, S., Bielow, C., Weisser, H., Aicheler, F., ... & Kohlbacher, O. (2016). OpenMS: a flexible open-source software platform for mass spectrometry data analysis. *Nature methods*, *13*(9), 741-748.

[15] Kong, A. T., Leprevost, F. V., Avtonomov, D. M., Mellacheruvu, D., & Nesvizhskii, A. I. (2017). MSFragger: ultrafast and comprehensive peptide identification in mass spectrometry–based proteomics. *Nature methods*, *14*(5), 513-520.

[16] Kou, Q., Xun, L., & Liu, X. (2016). TopPIC: a software tool for top-down mass spectrometry-based proteoform identification and characterization. *Bioinformatics*, *32*(22), 3495-3497.

[17] Beausoleil, S. A., Villén, J., Gerber, S. A., Rush, J., & Gygi, S. P. (2006). A probability-based approach for high-throughput protein phosphorylation analysis and site localization. *Nature biotechnology*, *24*(10), 1285-1292.

[18] Geiszler, D. J., Kong, A. T., Avtonomov, D. M., Yu, F., da Veiga Leprevost, F., & Nesvizhskii, A. I. (2021). PTM-Shepherd: analysis and summarization of post-translational and chemical modifications from open search results. *Molecular & Cellular Proteomics*, *20*.

[19] Käll, L., Canterbury, J. D., Weston, J., Noble, W. S., & MacCoss, M. J. (2007). Semi-supervised learning for peptide identification from shotgun proteomics datasets. *Nature methods*, *4*(11), 923-925.

[20] Yang, K. L., Yu, F., Teo, G. C., Li, K., Demichev, V., Ralser, M., & Nesvizhskii, A. I. (2023). MSBooster: improving peptide identification rates using deep learning-based features. *Nature Communications*, *14*(1), 4539.

[21] Elias, J. E., & Gygi, S. P. (2007). Target-decoy search strategy for increased confidence in large-scale protein identifications by mass spectrometry. *Nature methods*, *4*(3), 207-214.

[22] Zeng W F, Zhou X X, Willems S, et al. AlphaPeptDeep: a modular deep learning framework to predict peptide properties for proteomics[J]. Nature communications, 2022, 13(1): 7238.

[23] Larson, E. J., Pergande, M. R., Moss, M. E., Rossler, K. J., Wenger, R. K., Krichel, B., ... & Ge, Y. (2023). MASH Native: a unified solution for native top-down proteomics data processing. *Bioinformatics*, *39*(6), btad359.

[24] Liao, Y., Qian, R., Zhang, M., Sun, C., Wen, H., E, W., ... & Zhou, M. (2025). TDEase: An Open‐Source Data Visualization Software Framework for Targeted Proteoform Characterization by Top‐Down Proteomics. *Proteomics*, *25*(24), 22-29.

[25] Swearingen K E, Eng J K, Shteynberg D, et al. A Tandem Mass Spectrometry Sequence Database Search Method for Identification of O-Fucosylated Proteins by Mass Spectrometry[J]. Journal of proteome research, 2019, 18(2): 652-663.

[26] Park, J., Piehowski, P. D., Wilkins, C., Zhou, M., Mendoza, J., Fujimoto, G. M., ... & Kim, S. (2017). Informed-Proteomics: open-source software package for top-down proteomics. *Nature methods*, *14*(9), 909-914.

[27] Xu, A., Lin, B., Xue, B., Wang, B., Xu, B., Wu, B., ... & Wu, S. (2026). Deepseek-v4: Towards highly efficient million-token context intelligence. *arXiv preprint arXiv:2606.19348*.

[28] Adusumilli, R., & Mallick, P. (2017). Data conversion with ProteoWizard msConvert. In *Proteomics: methods and protocols* (pp. 339-368). New York, NY: Springer New York.

[29] Jeong, K., Kim, J., Gaikwad, M., Hidayah, S. N., Heikaus, L., Schlüter, H., & Kohlbacher, O. (2020). FLASHDeconv: ultrafast, high-quality feature deconvolution for top-down proteomics. *Cell Systems*, *10*(2), 213-218.

[30] Yu, F., Haynes, S. E., & Nesvizhskii, A. I. (2021). IonQuant enables accurate and sensitive label-free quantification with FDR-controlled match-between-runs. *Molecular & Cellular Proteomics*, *20*.

[31] Dupre M, Duchateau M, Malosse C, et al. Optimization of a top-down proteomics platform for closely related pathogenic bacterial discrimination[J]. Journal of proteome research, 2021, 20(1): 202-211.

[32] Melani R D, Gerbasi V R, Anderson L C, et al. The Blood Proteoform Atlas: A reference map of proteoforms in human hematopoietic cells[J]. Science, 2022, 375(6579): 411-418.

[33] Sadeghi S A, Chen W, Wang Q, et al. Pilot evaluation of the long-term reproducibility of capillary zone electrophoresis–tandem mass spectrometry for top-down proteomics of a complex proteome sample[J]. Journal of Proteome Research, 2024, 23(4): 1399.

[34] McCool E N, Xu T, Chen W, et al. Deep top-down proteomics revealed significant proteoform-level differences between metastatic and nonmetastatic colorectal cancer cells[J]. Science advances, 2022, 8(51): eabq6348.

[35] Jani V P, Yoo E J, Binek A, et al. Myocardial proteome in human heart failure with preserved ejection fraction[J]. Journal of the American Heart Association, 2025, 14(6): e038945.\

[36] Cai W, Zhang J, de Lange W J, et al. An unbiased proteomics method to assess the maturation of human pluripotent stem cell–derived cardiomyocytes[J]. Circulation research, 2019, 125(11): 936-953.

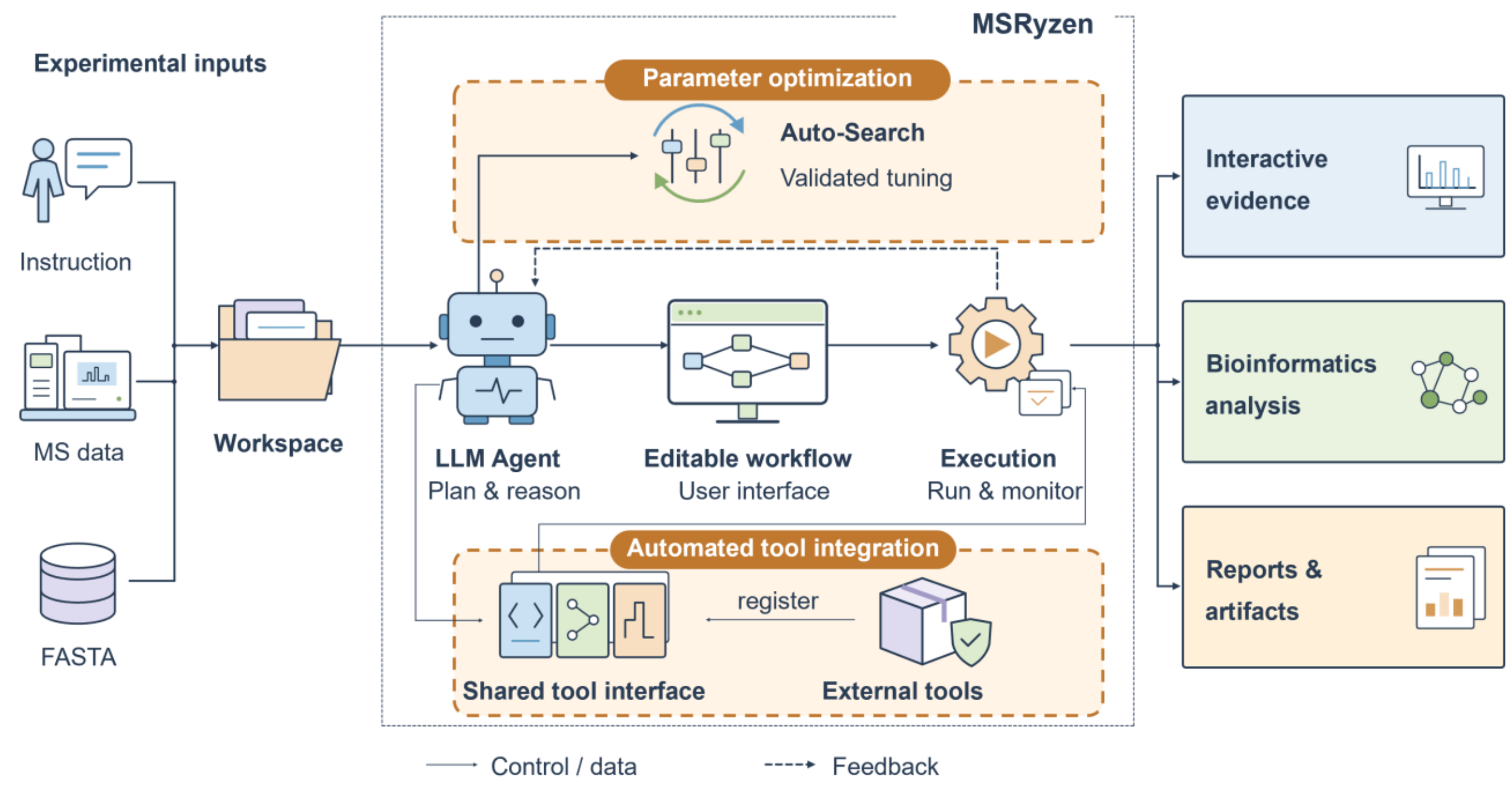

**Figure 1. MSRyzen platform architecture**. Natural-language prompts, MS data, and FASTA databases are processed via an LLM agent through an editable workflow directed acyclic graph (DAG). Database search results and quality metrics dynamically guide parameter optimization. Modular tool expansion is enabled via standardized adapters, and the user interface directly couples analytical outputs with spectral verification and downstream analyses.

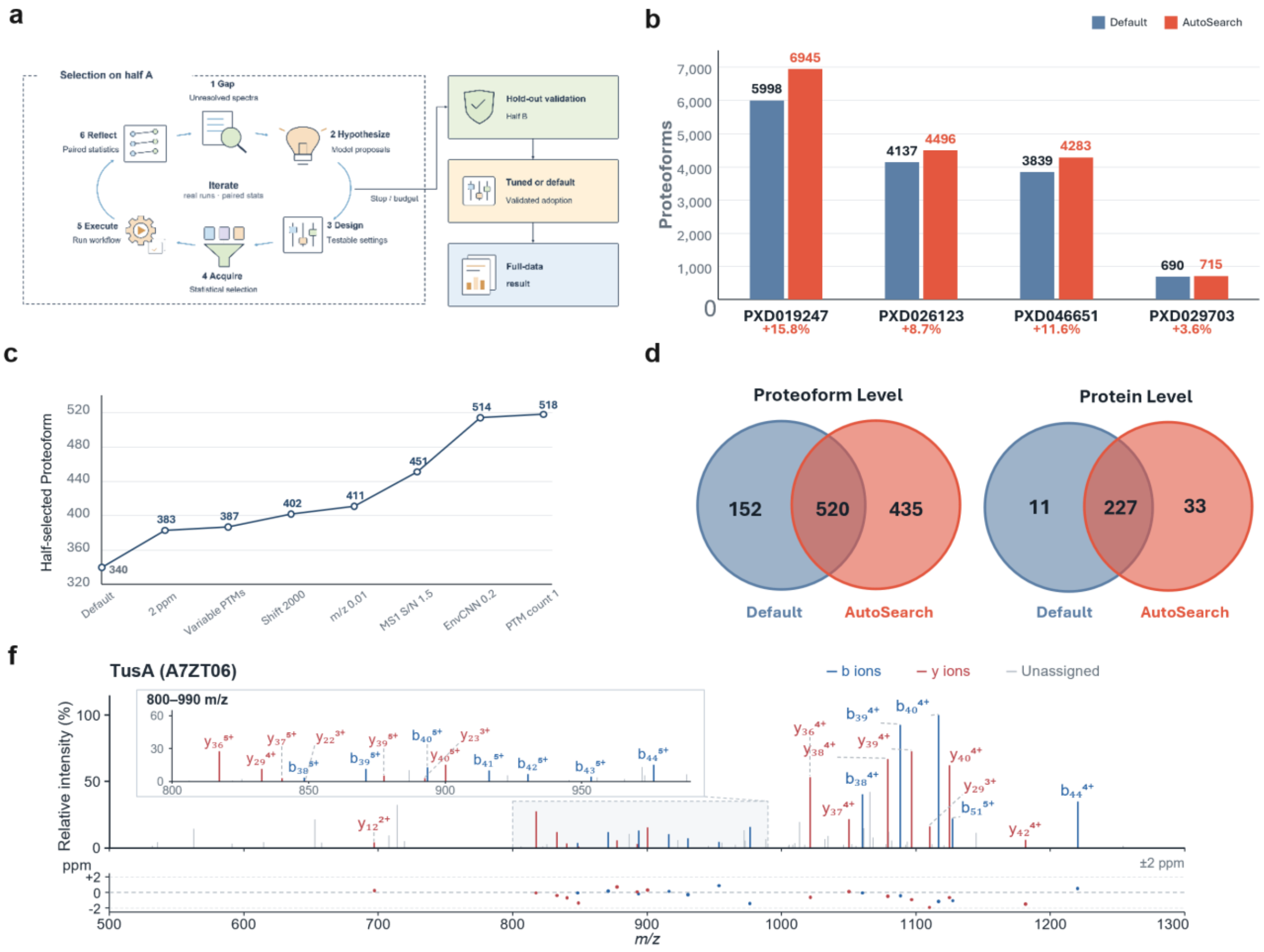

**Figure 2. Automatic parameter optimization for top-down proteomics.** a. Overview of the iterative optimization workflow, including candidate generation, parameter configuration, search execution, and result evaluation. b. Comparison of proteoform identifications obtained using default and optimized configurations across four public datasets at 1% FDR. c. Representative optimization trajectory across successive parameter configurations. d. Overlap of proteoform and protein identifications between default and optimized searches in a representative comparison. e. Comparison of modification classes and localization categories annotated under the two configurations. f. Fragment-ion annotations and mass-error distribution for a representative proteoform identification.

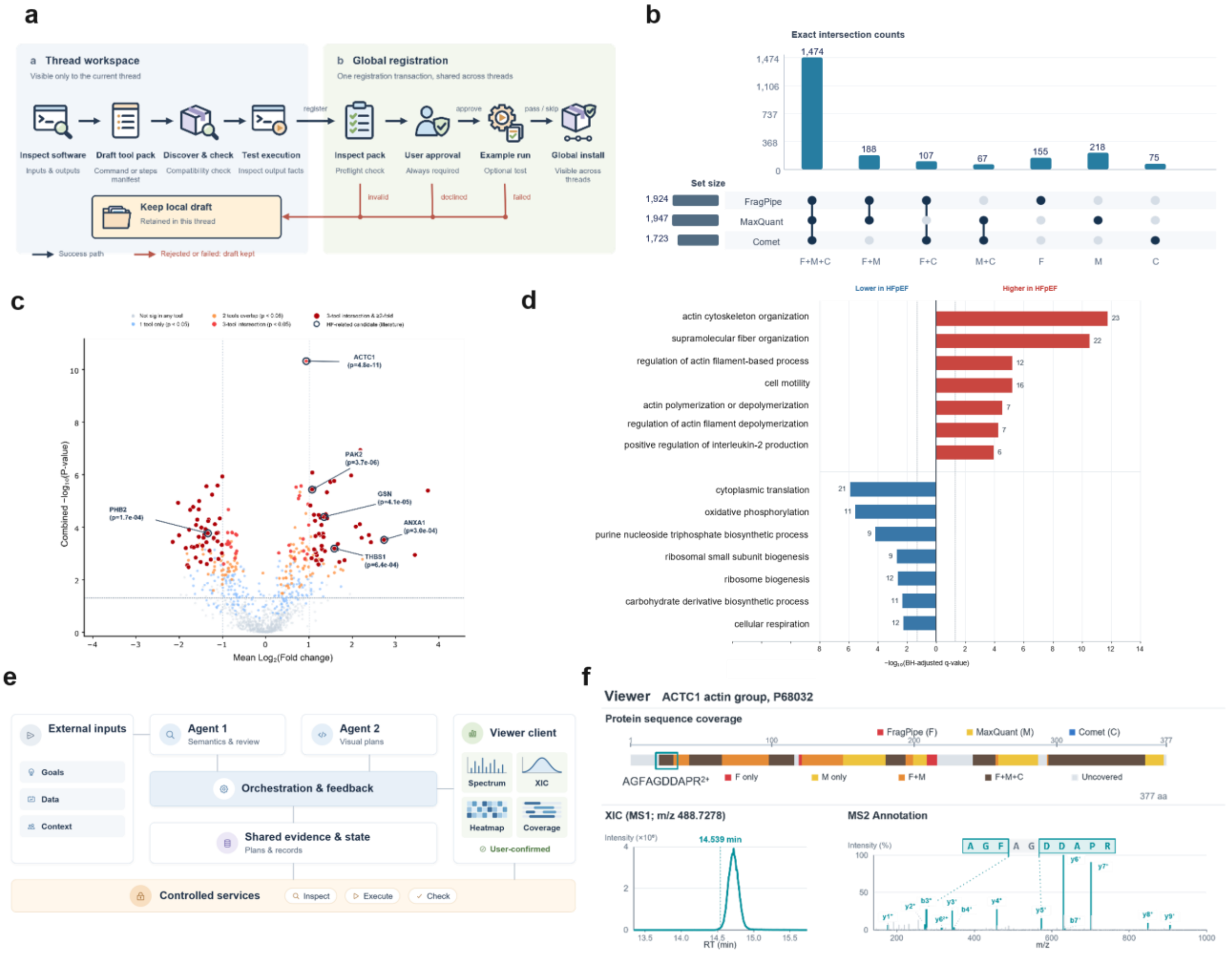

**Figure 3. Automated software integration and multi-tool proteomic analysis.** a. Schematic overview of local tool-pack development and user-confirmed global registration. b. UpSet plot illustrating protein-level overlap across FragPipe, MaxQuant, and Comet search outputs for dataset PXD060431. c. Integrated differential-abundance landscape comparing HFpEF (n = 10) against control (n = 10) cohorts based on mean log2 fold changes and Fisher-combined p-values. Color coding delineates proteins satisfying significance criteria in one, two, or all three search pipelines. d. Functional enrichment analysis of proteins exhibiting elevated (red) or diminished (blue) abundance in heart failure with preserved ejection fraction (HFpEF). e. Architecture of the dual-agent visualization framework coordinating evidence parsing and layout generation through a centralized task state. f. Inspectable spectral validation for the ACTC1 protein group, comprising cross-workflow sequence coverage maps, an extracted ion chromatogram, and the assigned MS2 fragmentation spectrum for doubly charged peptide AGFAGDDAPR (z = 2+).

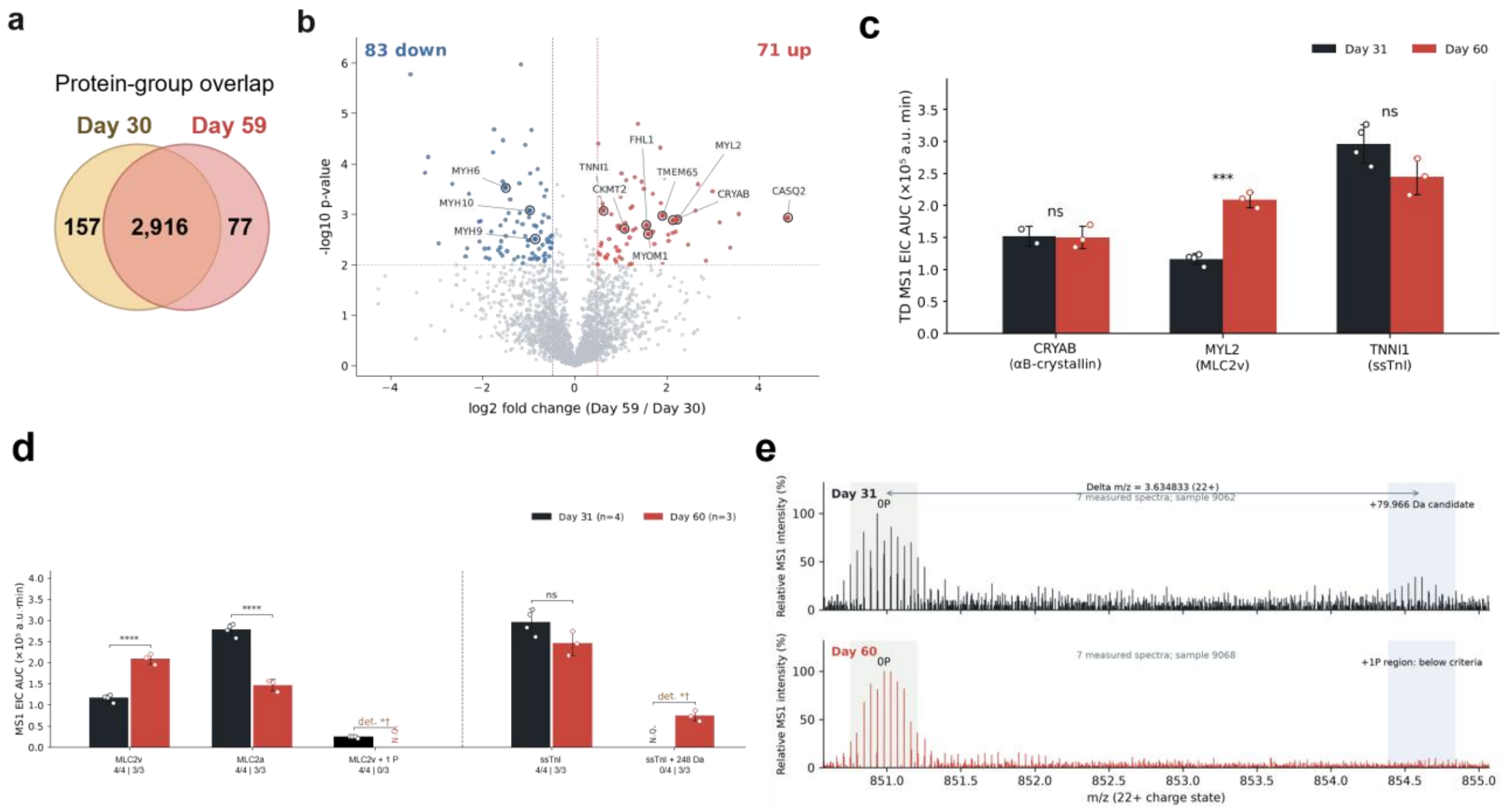

**Figure 4. Complementary bottom-up and intact-MS1 analysis of cardiomyocyte maturation.** a. Protein-group overlap between Day 30 and Day 59 bottom-up samples. b. Differential-abundance profiles of protein groups, with selected maturation-associated proteins labeled. c. Intact-MS1 quantification of CRYAB, MYL2, and TNNI1 in Day 31 and Day 60 samples. d. Quantification of selected intact-protein signals and candidate mass-shifted forms. Points represent individual runs (Day 31, n = 4; Day 60, n = 3). e. Representative MS1 spectra of the unshifted signal and the +79.966 Da candidate region at charge state 22+. The candidate region was below the detection criteria in the displayed Day 60 sample. Mass-shift assignments are based on MS1 evidence without MS2 confirmation. Bottom-up and intact-MS1 data represent distinct sample sets from MSV000082985. N.Q., not quantified.